\documentclass[twocolumn,10pt]{article}

\usepackage[T1]{fontenc}
\usepackage{lmodern}
\usepackage{textcomp}

\usepackage{dblfloatfix}
\usepackage{cancel}

\usepackage[english]{babel}

\usepackage[top=0.8in,bottom=1in,left=0.76in,right=0.76in]{geometry}
\usepackage{microtype}

\usepackage{amsmath,amssymb,amsfonts,mathtools}
\usepackage{bm}

\usepackage{graphicx}
\usepackage[dvipsnames]{xcolor}
\usepackage{authblk}
\usepackage[normalem]{ulem}

\usepackage{subcaption}
\usepackage{float}          
\usepackage{placeins}         
\usepackage[colorlinks=true,linkcolor=MidnightBlue,citecolor=ForestGreen,urlcolor=RoyalBlue]{hyperref}
\usepackage[capitalise,nameinlink,noabbrev]{cleveref} 
\newcommand{\R}{\mathbb{R}}

\title{
{ Super-Klein tunneling in 2D Lorentzian-type barriers in graphene}}
\author[1]{Alonso Contreras-Astorga}
\author[2]{Francisco Correa}
\author[3]{Luis Inzunza}
\author[4]{Vít Jakubský}
\author[5]{Raul Valencia-Torres}

\affil[1]{\textit{SECIHTI - Physics Department, Cinvestav, P.O. Box 14-740, 07000, Mexico City, Mexico.}}
\affil[2,3]{\textit{Departamento de F\'{i}sica, Universidad de Santiago de Chile, Avenida Victor Jara 3493, Estaci\'on Central, 9170124, Santiago, Chile.}}
\affil[4]{\textit{Nuclear Physics Institute of the Czech Academy of Science, Řež, Czech Republic}}
\affil[5]{\textit{Physics Department, Cinvestav, P.O. Box 14-740, 07000, Mexico City, Mexico.}}

\affil[ ]{\small \textit{Emails:} alonso.contreras@cinvestav.mx; francisco.correa.s@usach.cl; luis.inzunza@usach.cl; jakubsky@ujf.cas.cz; rvalenciat@ipn.mx}
\begin{document}

\twocolumn[
\begin{@twocolumnfalse}
\maketitle
\begin{abstract}
We introduce a two‑dimensional model of spin‑$1/2$ Dirac fermions in graphene subjected to a highly tunable electric field, which exhibits super‑Klein tunneling. The electric field can be continuously interpolated between two limiting configurations: a uniform electrostatic {Lorentzian} barrier with translational invariance and a chain of well‑separated electrostatic scatterers. We demonstrate that super-Klein tunneling arises naturally as a direct consequence of the intrinsic connection of the model to free-particle dynamics, a relation that is established through methods of supersymmetric quantum mechanics, which provide an elegant and analytically tractable framework. Besides the mentioned super-Klein tunneling, scale invariance of the model and invisibility of the potential for particles of specific energy are revealed, and possible routes toward experimental realization are discussed.\end{abstract}
\vspace{10mm}
\end{@twocolumnfalse}
]

\section{Introduction}


Super-Klein (or omnidirectional) tunneling is a phenomenon where a relativistic particle of specific energy passes through an electrostatic barrier without reflection, regardless of its incidence angle. This effect results from complex interference between reflected and transmitted waves at the barrier. The phenomenon has been explored in many physical scenarios \cite{shen}-\cite{LizarragaBrito2025}. Let us mention line-centered square lattices \cite{shen}, Dice lattices \cite{urban}, spin-1 systems with spin–orbit interaction \cite{Betancur1}, and transverse-polarized electromagnetic waves described by a spin-1 equation  \cite{louie}. It was also theoretically analyzed in time-periodic dice lattices \cite{Majari2022}, excited electromagnetic waves in bi-isotropic media \cite{Kim2020}, superlattices \cite{Wang2022}, and modeled in acoustic media \cite{Sirota2022}. Recently, super-Klein tunneling was observed in phononic crystals \cite{Zhu2023} and Lieb lattices \cite{Wu2024}.

Although its existence is often attributed to spin-1 systems with a flat band in the spectrum, super-Klein tunneling has also been shown to occur in spin-0 and spin-1/2 systems. In \cite{kim}, it was observed for spin-$0$ particles described by the Klein–Gordon equation, and further analyzed for spin-$1/2$ Dirac fermions in graphene \cite{Betancur2, SKT} as well as in the context of solitons \cite{SKT2}.

In all these scenarios, the effect arises in the presence of short-range interactions. Most studies considered an electric field in the form of a translationally invariant rectangular barrier, while in \cite{SKT} it was demonstrated for an exponentially decaying electric barrier lacking translational invariance.

In this Letter, we demonstrate that super-Klein tunneling occurs in spin-$1/2$ systems with a potential barrier exhibiting $x^{-2}$ asymptotics at large distances. We derive a one-parametric model of super-Klein tunneling where the electric field can vary from the Lorentzian-type barrier to highly structured comb of scatterers. 

\section{The model}
Similar to \cite{SKT}, we employ a combination of time-dependent SUSY transformations \cite{Samsonov} and a Wick rotation in the derivation of our model. The supersymmetric transformation produces a specific deformation of the potential in the evolution equation while preserving its solvability, mapping the solutions of the original system to those of the deformed one through the so-called intertwining operator. A subsequent Wick rotation then converts the $1+1$ evolution equation into a stationary equation of a planar system. The construction is described in more detail below.

Let us take the free Dirac particle in $1+1$ dimensions with mass $m$ as the initial system described by the following equation
\begin{equation}
{H_0 \psi=\left(i\partial_t-i\sigma_2\partial_z-m\sigma_3\right)\psi=0}.
\end{equation}
The equation can be solved in a straightforward manner by fixing the following two solutions,
\begin{eqnarray} 
{\phi}_1(z,t)&=&\left(i z \, e^{-i m t}  , \; \frac{i}{2m} \left( -e^{-i m t} + 2 \, \alpha \, e^{i m t}  \right) 
\right) ^{ T} \nonumber \\
{\phi}_2(z,t)&=&\sigma_1{\phi}_1(z,-t),\quad H_0\, {\phi}_{1,2}=0,\label{seed}
\end{eqnarray}
where $\alpha$ is a real constant. 
The two spinors (\ref{seed}) can be used to compose the matrix $U$,
\begin{equation}
U=({\phi_1},{\phi_2}),\qquad H_0\, U=0.
\end{equation}
Using $U$, we define the operators $H_1$ and $L$ through
\begin{eqnarray}
H_1&=&H_0-i\left[ \sigma_2,(\partial_zU)U^{-1} \right] \nonumber\\
&=&
 i\partial_t-i\sigma_2\partial_{z }-m\left(1-\frac{ \left( 1 - 2 \alpha \cos(2 m t) \right)}{F(-i z, t)}
\right)\,\sigma_3
\end{eqnarray}
and
\begin{eqnarray}
L&=&U\partial_{z}  U^{-1}\nonumber\\
&=&\partial_z + \frac{m}{F(-i z, t)} \left( 
    m\, z\, \sigma_{0} 
    + \left( \tfrac{1}{2} - \alpha \cos(2 m t) \right) \sigma_{1}\right. \nonumber\\
&&  \left.  - \alpha \sin(2 m t) \sigma_{2} 
\right), 
\end{eqnarray}
where  
\begin{equation}F(x,y)=m^{2}  x^{2} + \tfrac{1}{4} + \alpha\left( \alpha - \cos(2 m {y}) \right).
\end{equation}
These operators satisfy the following intertwining relation
\begin{equation}\label{intertwining}
LH_0=H_1L,
\end{equation}
see \cite{Samsonov} for more details. 
The operator $L$, commonly referred to as the supersymmetric or Darboux transformation, intertwines the operators $H_0$ and $H_1$. The latter corresponds to the deformed evolution equation $H_1\psi_1=0$, whose solutions can be generated by acting with $L$,
\begin{equation}\label{transform}
H_1\psi_1=0,\qquad \psi_1=L\psi_0,\qquad H_0\psi_0=0, 
\end{equation}
where $\psi_0$ is an arbitrary zero mode of $H_0$.

We now transform the $1+1$-dimensional system with a time-dependent potential into a stationary planar system\footnote{An alternative approach was employed in \cite{Louie}.} by the Wick rotation $z\rightarrow ix$, $t\rightarrow y$. Additionally, we multiply the equation $H_0\psi=0$  from the left by $\sigma_3$ and  subsequently perform a unitary transformation with the constant matrix $u=\sigma_3\,e^{i\pi/4\,\sigma_1}$. This yields
\begin{eqnarray}
h_0&=u\sigma_3H_0u^{-1}=&-i\sigma_1\partial_x-i\sigma_2\partial_y-m\,\sigma_0,\\
h_1&=u\sigma_3H_1u^{-1}=&-i\sigma_1\partial_x-i\sigma_2\partial_y-m {\sigma_0}\nonumber\\
&&+\frac{m \left( 1 - 2 \alpha \cos(2 m y) \right)}{F(x, y)}
\,\sigma_0,\end{eqnarray}
and
\begin{eqnarray}
\mathcal{L}&=u L u^{-1}=&-i\,\partial_x-\frac{m}{F(x, y)} \left(
    -i m  x \sigma_{0}
    + \right.
    \nonumber\\
    &&\left.\tfrac{1}{2} \left( 1 - 2 \alpha \cos(2 m y) \right) \sigma_{1}
    - \alpha\sin(2 m y) \sigma_{3}
\right).\nonumber  \\
\end{eqnarray}

As a result, the intertwining relation (\ref{intertwining}) between the three operators is modified and takes the form
\begin{equation} 
{\sigma_2\mathcal{L}\sigma_2}h_0=h_1\mathcal{L}.
\end{equation}
Due to the asymmetry of the intertwining operators on the left and right side of the equality, the relation provides the mapping between the zero modes of $h_0$ and those of $h_1$ only. 
The operator $h_1$ is Hermitian and represents the Hamiltonian of the planar system. Its potential term is regular provided that $F(x,y)$ has no zeros, which can be easily ensured by choosing $\alpha \neq 1/2$.

\begin{figure*}[!tb]
  \centering
  \begin{tabular}{ccc}
    \includegraphics[width=.29\textwidth]{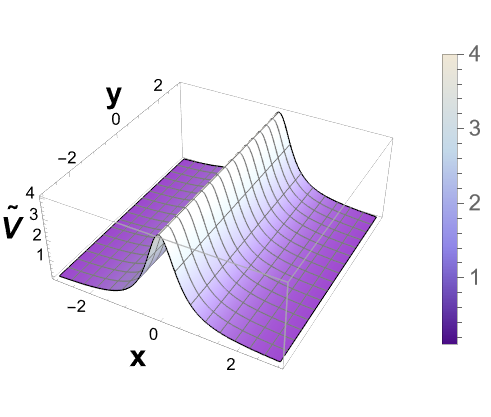} &
    \includegraphics[width=.29\textwidth]{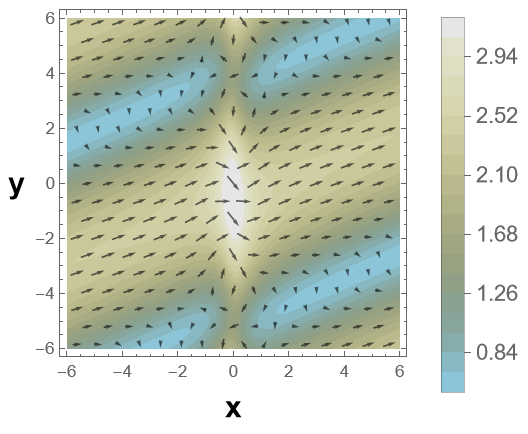} &
    \includegraphics[width=.29\textwidth]{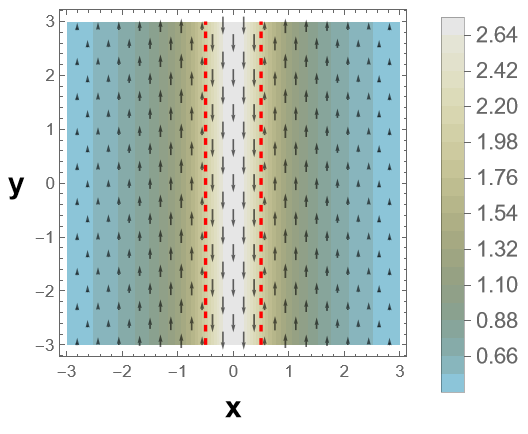} \\
    a)&b)&c)\\

    \includegraphics[width=.29\textwidth]{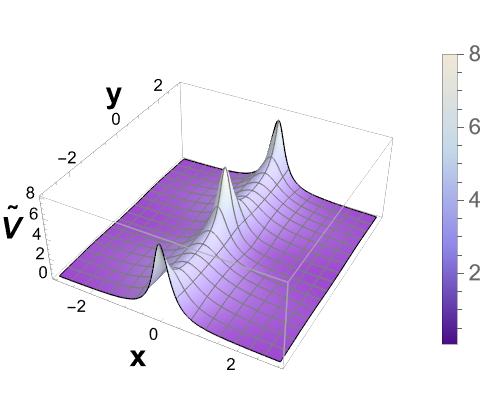} &
    \includegraphics[width=.29\textwidth]{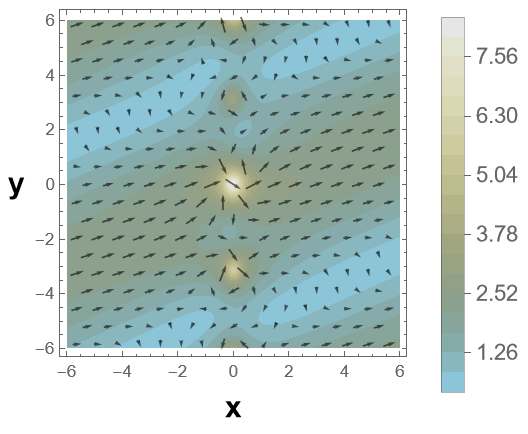} &
    \includegraphics[width=.29\textwidth]{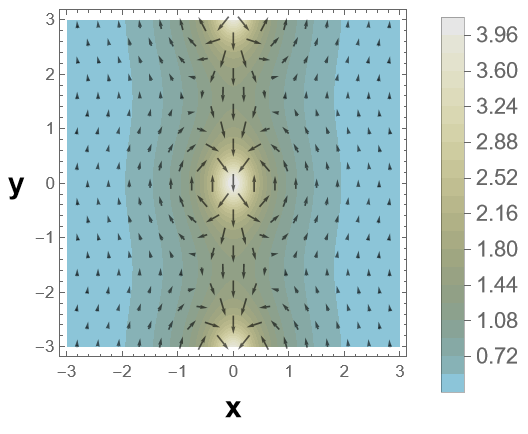}  \\
    d)&e)&f)\\

    \includegraphics[width=.29\textwidth]{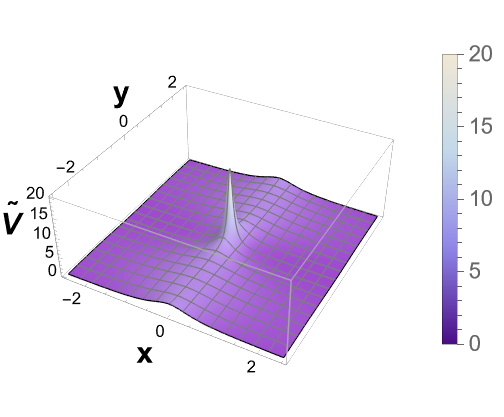} &
    \includegraphics[width=.29\textwidth]{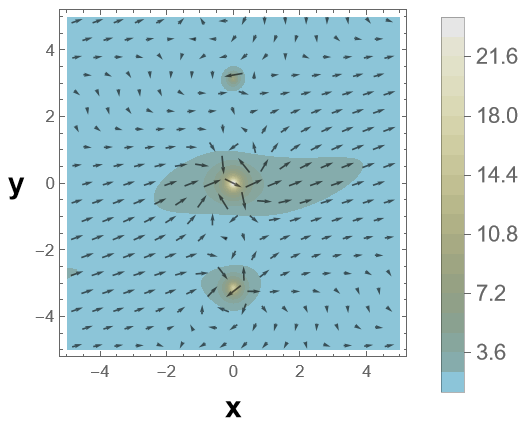} &
    \includegraphics[width=.29\textwidth]{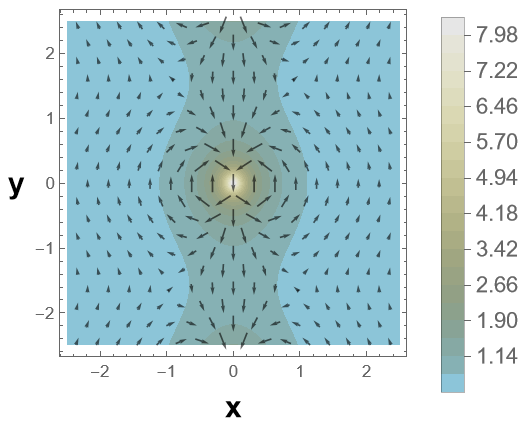}  \\
    g)&h)&i)\\

    \includegraphics[width=.29\textwidth]{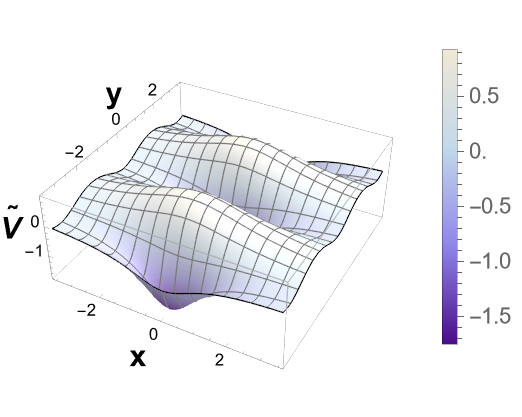} &
    \includegraphics[width=.29\textwidth]{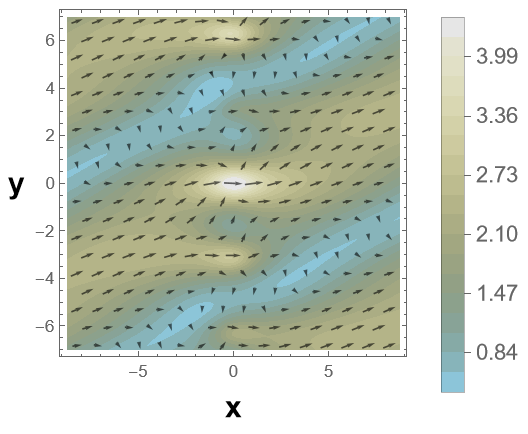} &
    \includegraphics[width=.29\textwidth]{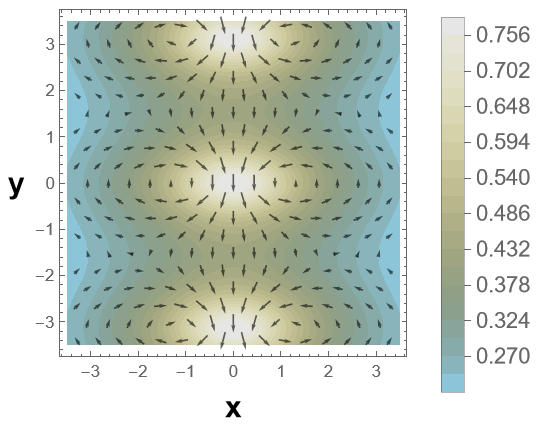}  \\
    j)&k)&l)
  \end{tabular}

  \caption{(left column) Potential $\tilde{V}(x,y)$.  (middle column) Probability density 
  of $\tilde{\psi}=\mathcal{L}\!\left(\psi_{0}+\psi_{\pi/4}\right)$
  together with its current $\vec{j}=(\widetilde{\psi})^{\dag} \vec{\sigma}\widetilde{\psi}$. The red lines in c) are borders of the areas where the probability current changes direction. (Right column) Probability density of localized state $\tilde{\psi}_\beta=\mathcal{L}\psi_\beta$ together with its current  $\vec{j}=(\widetilde{\psi}_{\beta})^{\dag} \vec{\sigma}\widetilde{\psi}_{\beta}$
  (right column).  We fixed a)-c) $\alpha=0$, d)-f) $\alpha=0.25$, g)-i) $\alpha=0.4$ and j)-l) 
  $\alpha=1.64$. In all figures, $m=1$ and $\beta=1$.}
  \label{fig1}
\end{figure*}

Extracting the constant part of the potential term in $h_1$, we obtain
\begin{equation}
\tilde{h}_1=-i\sigma_1\partial_x-i\sigma_2\partial_y+V(x,y),
\end{equation}
where (for $\alpha \in \mathbb{R}$)
\begin{align}
V(x,y)&=
\frac{m\left(1 - 2 \alpha \cos(2 m y) \right)}{m^{2}  x^{2} + \tfrac{1}{4} + \alpha \left( \alpha - \cos(2 m y) \right)}.\label{Vmodel}
\end{align}
The stationary equation associated with $\tilde h_1$ then reads
\begin{equation}\label{eqq}
\left(-i\sigma_1\partial_{{x}}-i\sigma_2\partial_{{y}}+V(x,y)\right)\tilde{\psi}(x,y)=m\tilde{\psi}(x,y),
\end{equation}
where $\tilde{\psi}(x,y)$ can be obtained from the zero modes of $h_0$, see (\ref{transform}). 
The potential $V(x,y)$ is periodic in $y$ and decays as $x^{-2}$ for large $|x|$,
\begin{equation}
V\left(x,y+\frac{\pi}{m}\right)=V(x,y),\qquad V(x,y)\sim
\frac{1}{m\,x^2}.
\end{equation}
The system possesses scale invariance, as under the rescaling
\begin{equation}
x= \tilde{x}/q,\qquad y= \tilde{y}/q, \qquad m={q\,\tilde{m}},
\end{equation} the equation (\ref{eqq}) remains the same. Therefore, we can set $m=1$ without loss of generality,
since other values of $m$ can be achieved by rescaling the coordinates. The potential is illustrated in Fig. \ref{fig1} for different values of the parameter $\alpha$. 
In the limit $\alpha \rightarrow 0$, we obtain a one-dimensional Lorentzian barrier
\begin{equation}
  V_L(x)= \lim_{\alpha \rightarrow 0} V(x,y)= 4 m \frac{(\frac{1}{2 m})^2}{x^2 + (\frac{1}{2 m})^2}
\label{L1D}
\end{equation}
centered at $x=0$, with a peak height of $4m$ and half-maximum points located at $x_{hm}=\pm 1/(2 m)$. In the subspace $p_y=0$, the Hamiltonian $\tilde{h}_1 \tilde{\psi}=E \tilde{\psi}$ with $\alpha=0$, is unitarily equivalent to the one-dimensional free particle energy operator \cite{Jakubsky1}. Therefore, its explicit eigenstates of arbitrary energy $E$ can be found as 
\begin{eqnarray}
    \tilde{\psi}_E=A_{+}f^{+}(x)\begin{pmatrix} 1 \\ -1 \end{pmatrix}
+iA_{-}f^{-}(x)\begin{pmatrix} 1 \\ 1 \end{pmatrix},
\end{eqnarray}
where $f^{\pm}(x)=\exp[\pm i \int^x (V_L(x')-E)dx' ]=\frac{(i\mp2mx)^2}{1+4m^2x^2}e^{\mp iEx}$.

Since $h_1$ was derived from the free-particle Hamiltonian, there is no back-scattering of Dirac fermions with energy $m$ on the potential barrier $V(x,y)$. Indeed, the scattering solutions of (\ref{eqq}) can be obtained from the plane waves 
\begin{equation} 
\psi_{\phi}=e^{i \mathbf{k}\cdot \mathbf{x}} (1,e^{i\phi})^{{T}}, \quad |\mathbf{k}|= m,
\end{equation}
where ${h}_0\psi_\phi=0$, {$\tilde{h}_1 (\mathcal{L}\psi_\phi) =m (\mathcal{L}\psi_{\phi})$}.
The scattering states of {$\tilde{h}_1$} with generic incidence angle $\phi$ can be generated from the free-particle plane waves $\psi_{\phi}$ upon acting with the operator $\mathcal{L}$. They have the following asymptotic form for $|x|\rightarrow\infty$,
\begin{equation}\label{Lasymptot}
\mathcal{L}\psi_{\phi}=\left(ik +O(x^{-1})\sigma_0+O\left(x^{-2}\right)\sigma_1+O\left(x^{-2}\right)\sigma_3\right)\psi_{\phi}.
\end{equation}
The wave function is not reflected as there is no component with the opposite momentum.  This implies that there is super-Klein tunneling in the system described by {$\tilde{h}_1$} for the particles of energy $E=m$.

Additionally, the wave function does not acquire any phase shift when going through the potential. Therefore, the potential $V(x,y)$ is completely invisible at this energy \cite{Mostafazadeh}. However, the invisibility is only asymptotic: the scattering state approaches a plane wave with corrections that are vanishing rather slowly as $x^{-1}$, see (\ref{Lasymptot}).
This is illustrated in Fig.\ref{fig1} where the probability density of a linear combination of asymptotic plane waves $\tilde{\psi}=\mathcal{L}\left(\psi_{0}+\frac{1}{2}\psi_{\pi/4}\right)$ is presented. The scattering states are affected by the presence of the barrier. For $\alpha\sim 1/2$ when the barrier is rather formed by a chain of sharply peaked scatterers, the probability density of the scattering states has sharp peaks at these scattering centers, see Fig.\ref{fig1}.

Besides the scattering states, there also exist localized states with normalizable localization along $x$-axis,  $\int_{-\infty}^{\infty}|\widetilde{\psi}_{\beta}|^2 dx<\infty$. We can derive such a localized state when applying the intertwining operator $\mathcal{L}$ on the following specific non-physical eigenstate of $h_0$ corresponding to zero energy,
\begin{eqnarray}
\psi_{\beta}= \frac{1}{\sqrt{2}}\left(
\begin{array}{c}
 -x e^{-i m y}-\frac{1}{2m}(2 \beta  e^{i m y}-e^{-i m y}) \\
 i xe^{-i m y}-\frac{i}{2m}(2 \beta  e^{i m y}-e^{-i m y}) \\
\end{array}
\right).
\end{eqnarray} 

The state $\tilde{\psi}_{\beta}=\mathcal{L}\psi_{\beta}$ is an eigenstate of $\tilde{h}_1$ with energy $m$, from which further bound states follow by exploiting the symmetries of $\tilde{h}_1$, e.g. $\sigma_1\tilde{\psi}_{\beta}(x,-y)$.  In Fig. \ref{fig1} we plot the non-normalized modulus
\begin{eqnarray}
|\widetilde{\psi}_{\beta}|^2= \frac{4 (\alpha -\beta )^2}{1+4 m^2 x^2+4 \alpha ^2-4 \alpha  \cos (2 m y)}.
\end{eqnarray}
Furthermore, due to the periodicity of the solutions, we can always interpret the system as an infinite cylinder defined by the restricted space $\Omega=\{(x,y)|x\in \R, y \in (0,\frac{\pi}{m})\}$, with the identification of the boundaries $y=0$ and $y=\frac{\pi}{m}$. In this case, the integration of the probability density can be performed exactly, and it yields 
\begin{equation}
\int_\Omega |\psi_\beta|^2d\Omega=4(\alpha-\beta)^2\frac{\pi K\left(-\frac{8 \alpha }{(1-2 \alpha )^2}\right)}{(1-2 \alpha ) m^2}\,,
\end{equation}
where $K(k)$ is the complete elliptic integral of the first kind.
Note that when $\alpha\rightarrow0$, this value reduces to $\frac{2\pi\beta^2}{m}$, obtaining in this way the bound state for one-dimensional Lorentzian barrier \eqref{L1D}.

It is worth noticing that the direction of the probability current $\vec{j}=(\widetilde{\psi}_{\beta})^{\dag} \vec{\sigma}\widetilde{\psi}_{\beta}$ changes the sign in the area of strong potential. In the limit case of a Lorentzian barrier (see \eqref{L1D}), the current in the $x$-direction vanishes everywhere; while the $y$ component has a change of sign exactly at the half-maximum point $x_{hm}= \pm 1/ (2m)$, where the potential $V(x_{hm},y)=2m$.  This behavior is generic for the current $j_y$ of the state $\tilde{\psi}_\beta$: for any fixed value of $y$, the position $x_0$ such that  condition $j_y=0$ holds (where there is a change of sign in the current):
\begin{equation}
    x_{0,\pm}=\pm \frac{1}{2m}\sqrt{4 \alpha ^2-4 \alpha  \cos (2 m y)+1}
\end{equation}
is equivalent to finding the half-maximum (or minimum) point of the potential $V(x_{0,\pm},y)/V(0,y)=1/2$.

\section{Experimental proposal at realistic scales}

To analyze the system at a realistic scale, we assume that $x$, $y$, and $m^{-1}$ have dimensions of length and multiply the equation $h\mathcal{L}\psi = m(\mathcal{L}\psi)$ by $\hbar v_f$. This identifies the physical energy as $\mathcal{E} = \hbar v_f m$, while the applied voltage is given by $U = \frac{\hbar v_f}{e}V(x, y)$. The maximal voltage is then obtained as $U_{\text{max}} = \frac{\hbar v_f}{e} V_{\text{max}} = 4\frac{\mathcal{E}}{e}\frac{1}{1-2\alpha}$, where the ratio $\frac{\mathcal{E}}{e}$ sets the voltage scale while $\alpha$ acts as a modulation parameter. As long as $\alpha \neq \frac{1}{2}$, the maximal voltage can be tuned according to the system scale determined by the parameter $m$. For example, if $\mathcal{E} = 0.1$ [eV], then $m = 0.152$ [nm$^{-1}$], resulting in a natural scale of $\pi/m = 20.6$ [nm], which exceeds the inter-atomic scale and thus validates the Dirac approximation. Furthermore, the effective channel length $1/m = 6.6$ [nm] defines a confining region accessible to a scanning tunnelling microscope (STM). In the following table we show how the maximum voltage behaves for different values of $\alpha$ between $0$ to $1$. \newpage
\begin{table}[h]
\centering
\begin{tabular}{|c|c|l|}
\hline
$\alpha$ & $U_{max}$ [V] & \textbf{Regime Description} \\ \hline
0.0 & 0.40 & Positive maximum potential (Barrier) \\ \hline
0.1 & 0.50 & Intensity increase \\ \hline
0.2 & 0.67 & Peak narrowing \\ \hline
0.3 & 1.00 & Near resonance \\ \hline
0.4 & 2.00 & Highly localized peak \\ \hline
0.6 & -2.00 & Inversion to deep potential well \\ \hline
0.7 & -1.00 & Potential well \\ \hline
0.8 & -0.67 & Depth decrease \\ \hline
0.9 & -0.50 & Stabilization \\ \hline
1.0 & -0.40 & Negative maximum well \\ \hline
\end{tabular}
\caption{Maximum values of the electric potential $U(x,y)$ in graphene evaluated at $x=0$ and $\cos(2my)=1$ for different values of the parameter $\alpha$.}
\label{tab:maximum_potential}
\end{table}

\begin{figure}[h]
  \centering
  \begin{subfigure}{1\linewidth}
    \centering
    \includegraphics[width=\linewidth, trim=2cm 4cm 1cm 6cm, clip]{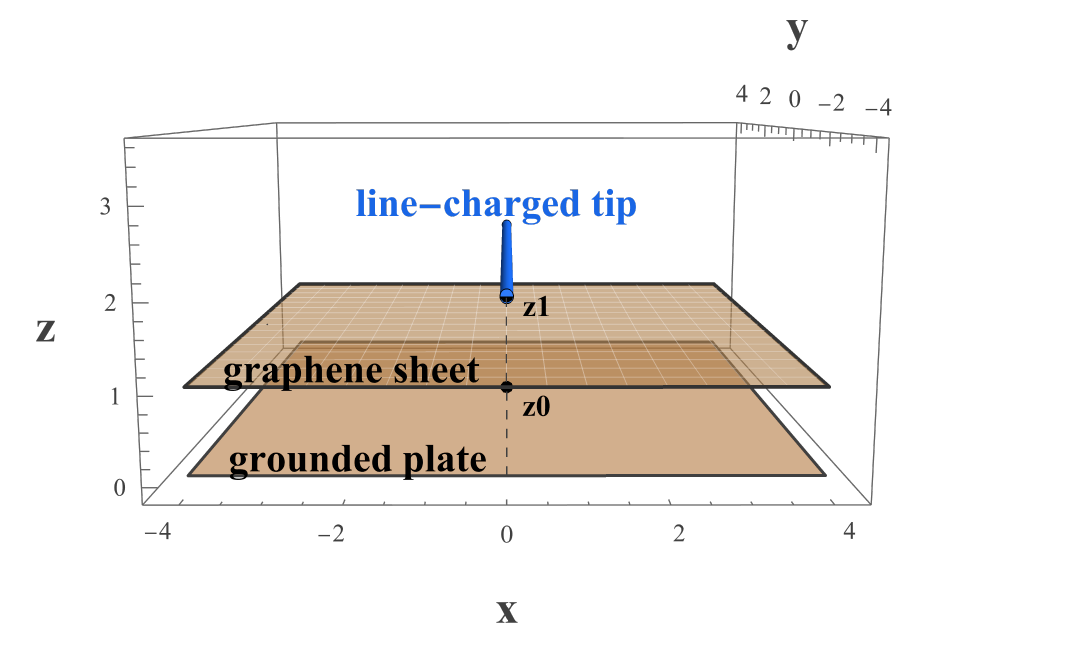}
    \caption{}
    \label{fig2a}
  \end{subfigure}

  \vspace{0.8em} 

  \begin{subfigure}{0.48\linewidth}
    \centering
    \includegraphics[width=\linewidth]{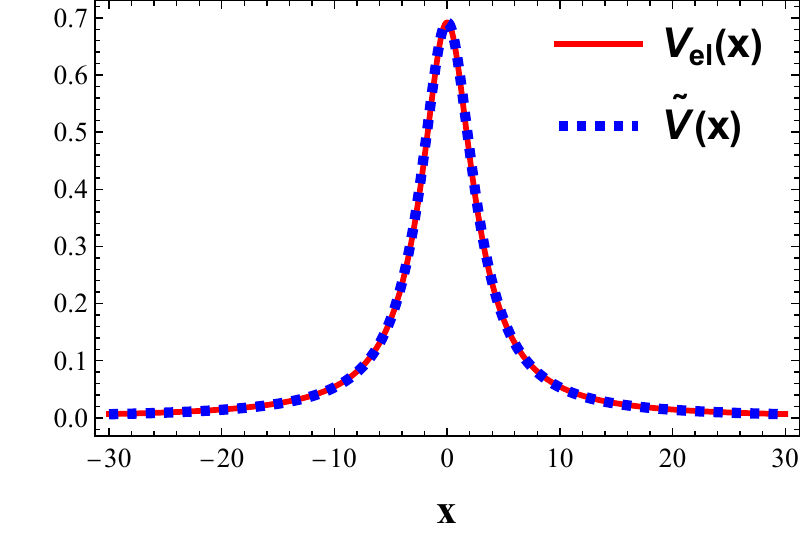}
    \caption{}
    \label{fig2b}
  \end{subfigure}
  \hfill
  \begin{subfigure}{0.48\linewidth}
    \centering
    \includegraphics[width=\linewidth]{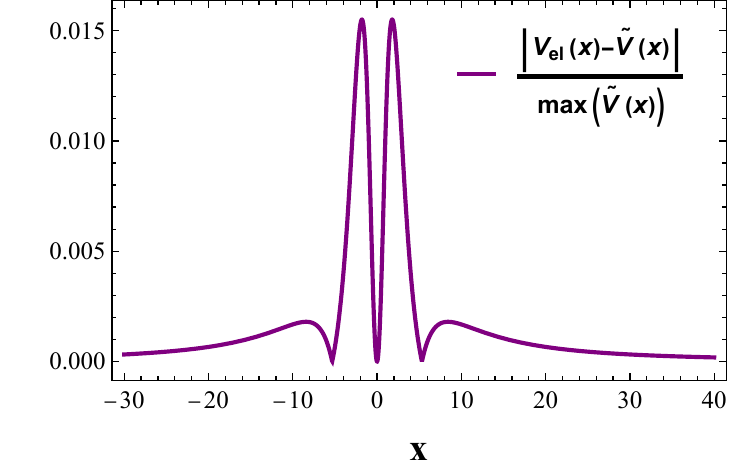}
    \caption{}
    \label{fig2c}
  \end{subfigure}

  \caption{a) Line-charged tip parallel to the graphene sheet and grounded plate. b) Potential $V_{el}(x,y)$ in (\ref{potentialel}) with charge density (\ref{tau}) (blue dashed) and $\tilde{V}$ for $\alpha=0$ (red dotted). c) The difference $\frac{|V_{el}-\tilde{V}|}{\mbox{max} \tilde{V}}$. We fixed $z_0=1$, $z_1=3$, $m=0.173$.}
  \label{fig2}
\end{figure}

The barrier represented by $V(x,y)$ can be constructed using the line-charged model of an STM tip placed in the proximity of a graphene sheet. Suppose that the graphene sheet lies in the plane $(x,y,z_0)$ with $z_0>0$, and that an infinite line with constant charge density $\tau$ passes through the points $(0,y,z_1)$, where $z_1>z_0$, see Fig.~\ref{fig2a} for illustration. The plane $(x,y,0)$ is grounded, i.e. the electrostatic potential generated by the charged line must vanish there, and the electric potential on the graphene sheet can then be obtained using the method of mirror charges, yielding
\begin{equation}\label{potentialel}V_{el}(x)=\tau
\ln\left(
\frac{x^{2} + (z_{0}+z_{1})^{2}}
     {x^{2} + (z_{0}-z_{1})^{2}}
\right),\end{equation}
To reproduce the barrier profile (\ref{Vmodel}), we fix the charge density of the line as 
\begin{equation}
\tau=\frac{4 m}{\ln\!\left(\dfrac{(z_{0}+z_{1})^{2}}{(z_{0}-z_{1})^{2}}\right)}
\;.\label{tau}
\end{equation}
With this choice, the electric potential of the line-charged tip (\ref{potentialel}) is accurately approximated by the barrier (\ref{Vmodel}) for $\alpha=0$, as illustrated in Fig.~\ref{fig2b} and Fig.~\ref{fig2c}.

\section{Summary}

We derived a one-parameter family of models describing pseudo-spin-$\tfrac{1}{2}$ Dirac fermions in the presence of an electric potential (\ref{Vmodel}) that has an asymptotic $1/x^{2}$ dependence. Depending on the value of the parameter, the barrier is either translationally invariant along the $y$ axis ($\alpha = 0$) or forms a $y$-periodic chain of scatterers. The construction is based on a time-dependent supersymmetric transformation that makes the stationary equation analytically solvable for a specific value of the energy.  
It allowed us to find the scattering states and show that the system possesses super-Klein tunneling for the particles with $E=m$.

Contrary to the electrostatic barrier found in \cite{SKT}, 
the transmitted plane waves acquire no phase shift, see (\ref{Lasymptot}). 
Consequently, the barrier is completely undetectable at $E = m$, and it is effectively invisible to these scattering states. Comparing the experimental feasibility, the electric field behaves like $1/x^2$ that is easier to prepare in the experiments than the exponentially vanishing one \cite{Downing1},\cite{Downing2}, \cite{Szafran}. We demonstrated that the electric field in the form of the Lorentzian barrier approximates very well the field of a prolonged, line-charged STM tip placed in the proximity of the graphene sheet.

It is worth noting that Dirac fermions in electric fields with $1/r^2$ asymptotic behavior have been studied, e.g., in \cite{Downing1}-\cite{Szafran}, where confinement of zero-energy charge carriers by the electric field was observed. In \cite{LorentzianDots}, the supersymmetric framework underlying the existence of zero modes in generalized Lorentzian quantum dots was discussed.

\section*{Acknowledgements}
{ACA acknowledges Se\-cretar\'ia de Ciencia, Humanidades, Tecnolog\'ia e Innovaci\'on (SECIHTI - M\'exico) support under the grant FORDECYT-PRONACES/61533/2020. F. C. was supported by Fondecyt Grants No. 1252036, No. 13250014 and by USACH project 042531CS$\_$Ayudante. L. I was supported by USACH project 042531CS$\_$Ayudante. VJ acknowledges the assistance provided by the Advanced Multiscale Materials for Key Enabling Technologies project, supported by the Ministry of Education, Youth, and Sports of the Czech Republic. Project No. CZ.02.01.01/00/22 008/0004558, Co-funded by the European Union.”AMULET project.}

\end{document}